\documentclass[12pt,preprint,showkeywords]{iopart}

\usepackage{graphicx}
\usepackage{dcolumn}
\usepackage{iopams}
\usepackage{amssymb}
\usepackage{latexsym}
\usepackage{color}

\begin{document}
\title[Cross-correlation of long-range correlated series]{Cross-correlation of long-range correlated series}
\author{Sergio Arianos and Anna Carbone}
\address{Physics Department, Politecnico di Torino,\\ Corso Duca
degli Abruzzi 24, 10129 Torino, Italy}
\ead{sergio.arianos@polito.it,anna.carbone@polito.it}
\date{\today}

\begin{abstract}
A  method  for estimating the cross-correlation
$C_{xy}(\tau)$ of long-range correlated series $x(t)$ and $y(t)$,  at
varying lags $\tau$ and scales $n$, is proposed. For fractional Brownian
motions with Hurst exponents $H_1$ and $H_2$, the asymptotic
expression of $C_{xy}(\tau)$ depends only on the lag $\tau$ (wide-sense stationarity) and scales
as a power of $n$  with exponent ${H_1+H_2}$  for $\tau\rightarrow 0$. The method is
illustrated
 on (\emph{i}) financial series, to show the leverage effect; (\emph{ii}) genomic
sequences, to estimate the correlations between structural parameters along
the chromosomes.\par\noindent
\end{abstract}

\maketitle

\section{Introduction and overview}
Interdependent behaviour and causality in coupled complex systems
continue to attract considerable interest in fields as diverse as
solid state science, biology, physiology, climatology
\cite{Rosenblum,Zhou,Oberholzer,Dhamala,Verdes,Palus,Kreuz,Du}.
Coupling and synchronization effects have been observed for
example in cardiorespiratory interactions, in neural signals, in
glacial variability and in Milankovitch forcing
\cite{Tass,Huybers,Ashkenazy}. In finance, the \emph{leverage
effect}  quantifies the
cause-effect relation between return $r(t)$ and volatility
$\sigma_T(t+\tau)$  and eventually financial risk estimates \cite{Black,Schwert,Haugen,Glosten,Wu1,Figlewski,Bouchaud,Perello,Qiu,Ahlgren,Varga,Montero}. In
DNA sequences, causal connections among structural and compositional properties such
as intrinsic curvature, flexibility, stacking energy, nucleotide
composition are sought to unravel the mechanisms underlying
biological processes in cells \cite{Moukhtar,Allen,Pedersen}.
\par Many issues still remain unsolved mostly due to problems with the accuracy and resolution of  coupling estimates
in long-range correlated signals. Such signals do not show the
wide-sense-stationarity
 needed to yield statistically meaningful information when  cross-correlations and cross-spectra are estimated.
  In \cite{Jun,Podobnik}, a function
$F_{xy}(n)$, based on the
detrended fluctuation analysis - a measure of autocorrelation of a series at different scales $n$ - has been proposed to
estimate the cross-correlation of two
series $x(t)$ and $y(t)$. However, the function
$F_{xy}(n)$  is independent of the lag $\tau$, since it is a straightforward generalization of the detrended fluctuation analysis, which is a
\emph{positive-defined} measure of autocorrelation for long-range
correlated series. Therefore, $F_{xy}(n)$ holds only for $\tau=0$. Different from
autocorrelation, the cross-correlation of two long-range
correlated signals is a \emph{non-positive-defined function of $\tau$}, since the coupling could be delayed  and
have any sign.
\par In this work, a method to estimate the cross-correlation
function $C_{xy}(\tau)$ between two long-range correlated signals
at different scales $n$ and lags $\tau$ is developed. The
asymptotic expression of $C_{xy}(\tau)$ is worked out for fractional Brownian motions $B_H(t)$, $H$ being the Hurst
exponent, whose interest follows from their widespread use for  modeling
long-range correlated processes in different areas \cite{Mandelbrot}. Finally, the method is used to investigate the coupling
between
(\emph{i}) returns and volatility of the DAX stock index and (\emph{ii})
structural properties, such as deformability,
stacking energy, position preference and propeller twist,  of
the Escherichia Coli chromosome.

\par
  The
proposed method  operates:  (i)  on the integrated
rather than on the increment series, thus yielding the
cross-correlation at varying windows $n$, as opposed to the
standard cross-correlation;  (ii)   as a sliding
product of two series,  thus yielding the cross-correlation as a
function of the lag $\tau$, as opposed to the method proposed in \cite{Jun,Podobnik}. The features (i)  and (ii)  imply higher
accuracy,  $n$-windowed resolution while capturing the cross-correlation at
varying lags $\tau$.

\section{Method}
The \emph{cross-correlation} $C_{xy}(t,\,\tau)$ of two
nonstationary stochastic processes $x(t)$ and $y(t)$ is defined as:
\begin{equation}\label{crosscovariance}
C_{xy}(t,\tau)\equiv
\Big\langle[x(t)-\eta_x(t)][y^\ast(t+\tau)-\eta_y^\ast(t+\tau)]\Big\rangle
\end{equation}
where $\eta_x(t)$ and $\eta_y^\ast(t+\tau)$ indicate time-dependent
means of $x(t)$ and
 $y^\ast(t+\tau)$, the symbol
 $\ast$ indicates the complex conjugate and the brackets $<>$ indicate the ensemble average over
 the joint domain of $x(t)$ and
 $y^\ast(t+\tau)$. This relationship holds for space dependent sequences,
  as for example the chromosomes, by replacing time with space coordinate.
  Eq.~(\ref{crosscovariance}) yields sound information provided the two quantities in
square parentheses are jointly stationary and thus
$C_{xy}(t,\,\tau)\equiv C_{xy}(\tau)$ is a function only of the lag $\tau$.

In this work, we propose to estimate the
cross-correlation  of two nonstationary signals by choosing for
 $\eta_x(t)$ and $\eta_y^\ast(t+\tau)$ in Eq.~(\ref{crosscovariance}), respectively the functions:
\begin{equation} \label{xtil}
\widetilde{x}_n(t)  =  \frac{1}{n}\sum_{k=0}^nx(t-k)
\end{equation}
and
\begin{equation} \label{ytil}
\widetilde{y}_n^*(t+\tau) =  \frac{1}{n}\sum_{k=0}^ny^*(t+\tau-k)
\end{equation}
\subsection{Wide-sense stationarity}
The wide-sense stationarity of Eq.~(\ref{crosscovariance}) can be demonstrated for fractional Brownian motions. By taking $x(t)=B_{H_1}(t)$, $y(t)=B_{H_2}(t)$,
$\eta_x(t)$ and $\eta_y^\ast(t+\tau)$ calculated according to
Eqs.~(\ref{xtil},\ref{ytil}), $C_{xy}(t,\,\tau)$
writes:
\begin{eqnarray}
\label{dcaB0} C_{xy}(t,\,\tau)
=\Big\langle\big[B_{H_1}(t)-\widetilde{B}_{H_1}(t)\big]\big[B_{H_2}^*(t+\tau)-\widetilde{B}_{H_2}^*(t+\tau)\big]\Big\rangle
\;\;\;.
\end{eqnarray}
\noindent When writing  $x(t)=B_{H_1}(t)$ and $y(t)=B_{H_2}(t)$, we assume the same underlying generating noise $dB(t)$ to produce a sample of $x$ and $y$. Eq.~(\ref{dcaB0}) is
calculated in the limit of large $n$ (calculation details are reported in
the Appendix). One obtains:

\begin{eqnarray}
\label{theta} C_{xy}(\hat{\tau}) & =
n^{H_1+H_2}D_{H_1,\,H_2}\Big[-\hat{\tau}^{H_1+H_2}\nonumber\\
&+\frac{(1+\hat{\tau})^{1+H_1+H_2}+(1-\hat{\tau})^{1+H_1+H_2}}{1+H_1+H_2}\nonumber\\
&-\frac{(1-\hat{\tau})^{2+H_1+H_2}-2\hat{\tau}^{2+H_1+H_2}+(1+\hat{\tau})^{2+H_1+H_2}}{(1+H_1+H_2)(2+H_1+H_2)}
\Big]\;\;\;,
\end{eqnarray}
\noindent
where $\hat{\tau} = \tau/n $ is the \emph{scaled lag} and
$D_{H_1,\,H_2}$ is defined in
the Appendix.  Eq.~(\ref{theta})
is independent of $t$, since the terms in square parentheses
depend only on $\hat{\tau} = \tau/n$, and  thus
Eq.~(\ref{crosscovariance}) is made wide-sense stationary. It is worthy of
note that, in Eq.~(\ref{theta}), the coupling between $B_{H_1}(t)$ and $B_{H_2}(t)$ reduces to the sum of the exponents $H_1 + H_2$.
Eq.~(\ref{theta}), for $\tau=0$,
reduces to:
\begin{eqnarray}
\label{zero} C_{xy}(0) \propto n^{H_1+H_2}  \;\;\;,
\end{eqnarray}
indicating that the coupling between $B_{H_1}(t)$ and $B_{H_2}(t)$
 scales as the product of  $n^{H_1}$ and $n^{H_2}$. The property of the variance of fractional Brownian motion $B_{H}(t)$ to
scale as $n^{2H}$ is recovered from the
Eq.~(\ref{zero}) for $x=y$ and $H_1=H_2=H$, i.e.:
\begin{eqnarray}
\label{auto} C_{xx}(0) \propto  n^{2H} \;\;\;.
\end{eqnarray}
Eq.~(\ref{auto}) has been studied in
\cite{Carbone1,Carbone2,Carbone3,Carbone4,Carbone5}. \\ \noindent

\section{Examples}
  \subsection{Financial series} 
\begin{figure}
\begin{center}
\includegraphics[width=8cm]{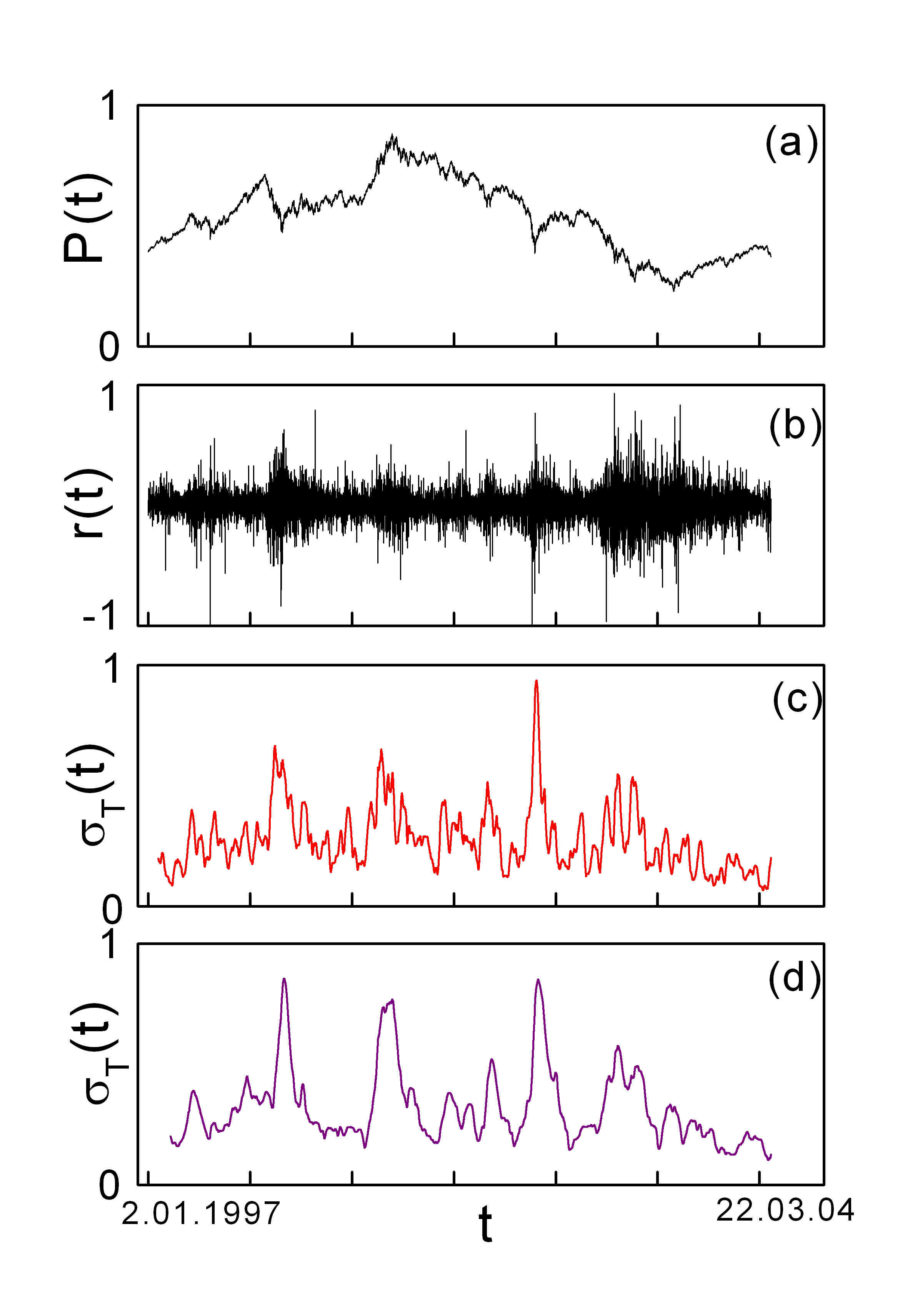}
\caption{\label{fig:figure1}   DAX stock index: (a)
prices; (b) returns with  $t'=1h$;  (c) volatility
  with $T=300h$; (d) volatility with $T=660h$.}
\end{center}
\end{figure}

The leverage effect is a \emph{stylized fact} of finance. The level of volatility is related to  whether returns are negative or positive.  Volatility rises when  a stock's price drops and falls when the stock goes up \cite{Black}. Furthermore, the impact of negative returns on volatility seems much stronger than the impact of positive returns  (\emph{down market effect}) \cite{Wu1,Figlewski}.
To illustrate these
  effects, we analyze the correlation between returns and volatility  of the
DAX stock index $P(t)$, sampled every minute from
2-Jan-1997 to 22-Mar-2004, shown in Fig.~\ref{fig:figure1}~(a).
The
returns and volatility are defined respectively as:
$r(t)=\ln P(t+t')- \ln P(t)$
and
 $ \sigma_T(t)=\sqrt{{\sum_{t=1}^{T}\big[r(t)-\overline{r(t)}_T\big]^2}/{(T-1)}}
  \;.$

\begin{figure}
\begin{center}
\includegraphics[width=8cm]{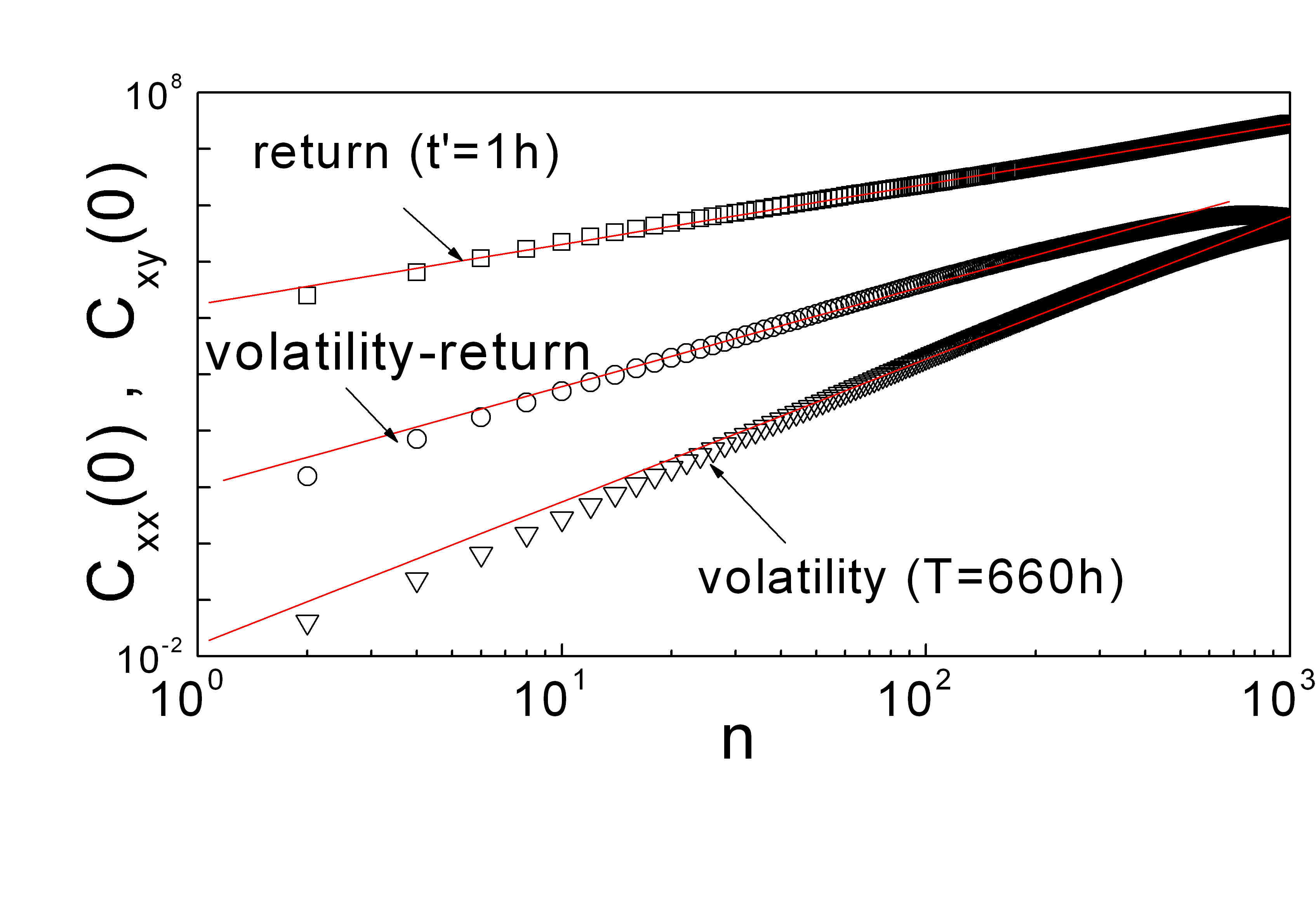}
\caption{\label{fig:figure2}  Log-log plot of
$C_{xx}(0)$ for the DAX return (squares) and volatility (triangles) and of $C_{xy}(0)$ with $x=r(t)$ and $y=\sigma_T(t)$ (circles). Red lines are linear fits. The power-law behaviour is consistent with
Eqs.~(\ref{zero},\ref{auto}).}
\end{center}
\end{figure}

\begin{figure}
\begin{center}
\includegraphics[width=8cm]{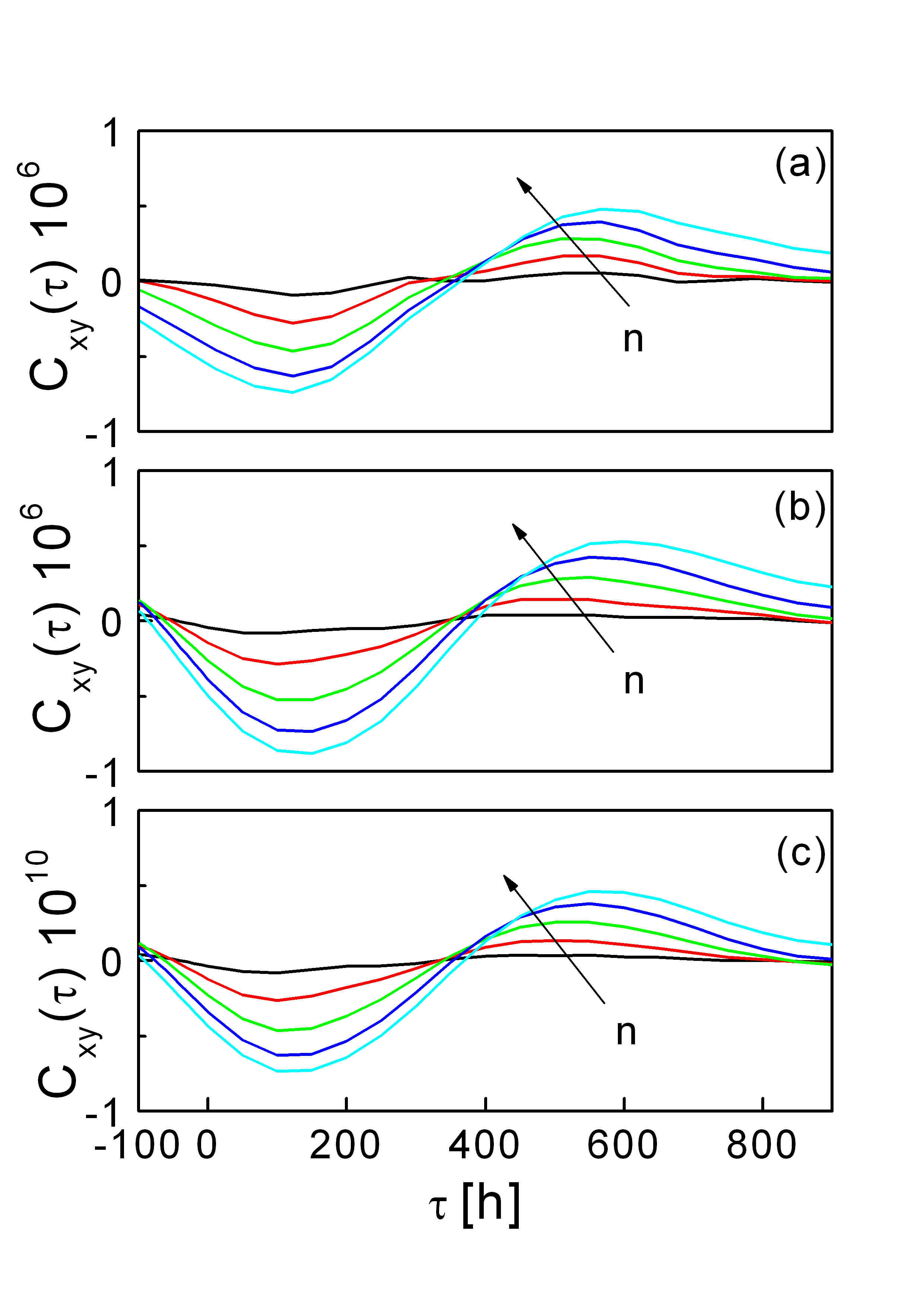}
\caption{\label{fig:figure3}   Cross-correlation
$C_{xy}(\tau)$ with  $x=r(t)$ and $y=\sigma_T(t)$   with (a)
$T=300h$
 and (b) $T=660h$;  (c) with $x=r(t)$ and $y=\sigma_T(t)^2$ with $T=660h$. $n$ ranges from 100 to 500 with step 100.}
\end{center}
\end{figure}

\noindent Fig.~\ref{fig:figure1}~(b) shows the returns for $t'=1 h$. The volatility series are
 shown   in Figs.~\ref{fig:figure1}~(c,d) respectively for
$T=300h$ and $T=660h$.
The Hurst exponents, calculated by the slope of the log-log plot
of Eq.~(\ref{auto}) as a function of $n$, are $H=0.50$ (return), $H=0.77$ (volatility $T=300h$)
 and $H=0.80$ (volatility $T=660h$). Fig.~\ref{fig:figure2} shows the log-log plots of
 $C_{xx}(0)$  for the returns
 (squares) and volatility with $T=660$ (triangles). The scaling-law exhibited by the DAX series guarantees that its behaviour is a fractional Brownian motion. The function $C_{xy}(0)$
 with $x=r(t)$ and
$y=\sigma_T(t)$ with $T=660h$ is also plotted at varying $n$ in Fig.~\ref{fig:figure2}  (circles). From the slope of the log-log plot of
$C_{xy}(0)$ vs $n$, one obtains $H=0.65$, i.e. the average between $H_1$ and
$H_2$ as expected from Eq.~(\ref{zero}). \\
\noindent
 Next, the cross-correlation is considered
 as a function of $\tau$. The plots of
$C_{xy}(\tau)$ for  $x=r(t)$ and $y=\sigma_T(t)$ with
$T=300h$ and $T=660h$ are shown respectively in
Fig.~\ref{fig:figure3}~(a,b) at different windows $n$.

\begin{figure}
\begin{center}
\includegraphics[width=8cm]{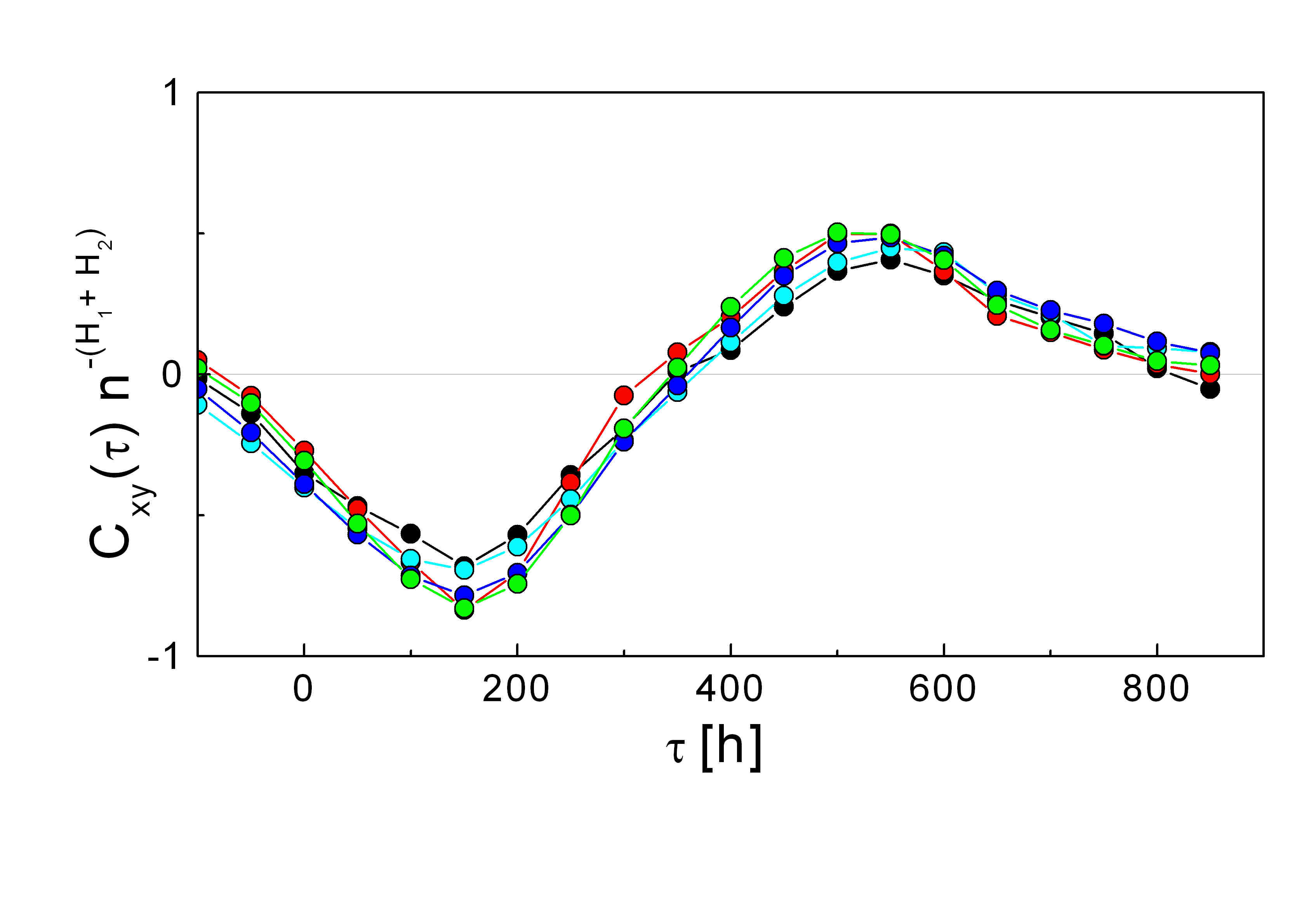}
\caption{\label{fig:figure4}   Plot of the function
$C_{xy}(\tau) n^{-(H_1+H_2)}$ with  $x=r(t)$ and $y=\sigma_T(t)$   with
$T=300h$.  $H_1=0.5$ and $H_2=0.77$ $n$ ranges from 100 to 500 with step 100. One can note that the five curves collapse, within the numerical errors of the parameters entering the auto- and cross-corerlation functions. This is in accord with the invariance of the  product $C_{xy}(\tau) n^{-(H_1+H_2)}$ with the window $n$.}
\end{center}
\end{figure}

\begin{figure}
\begin{center}
\includegraphics[width=8cm]{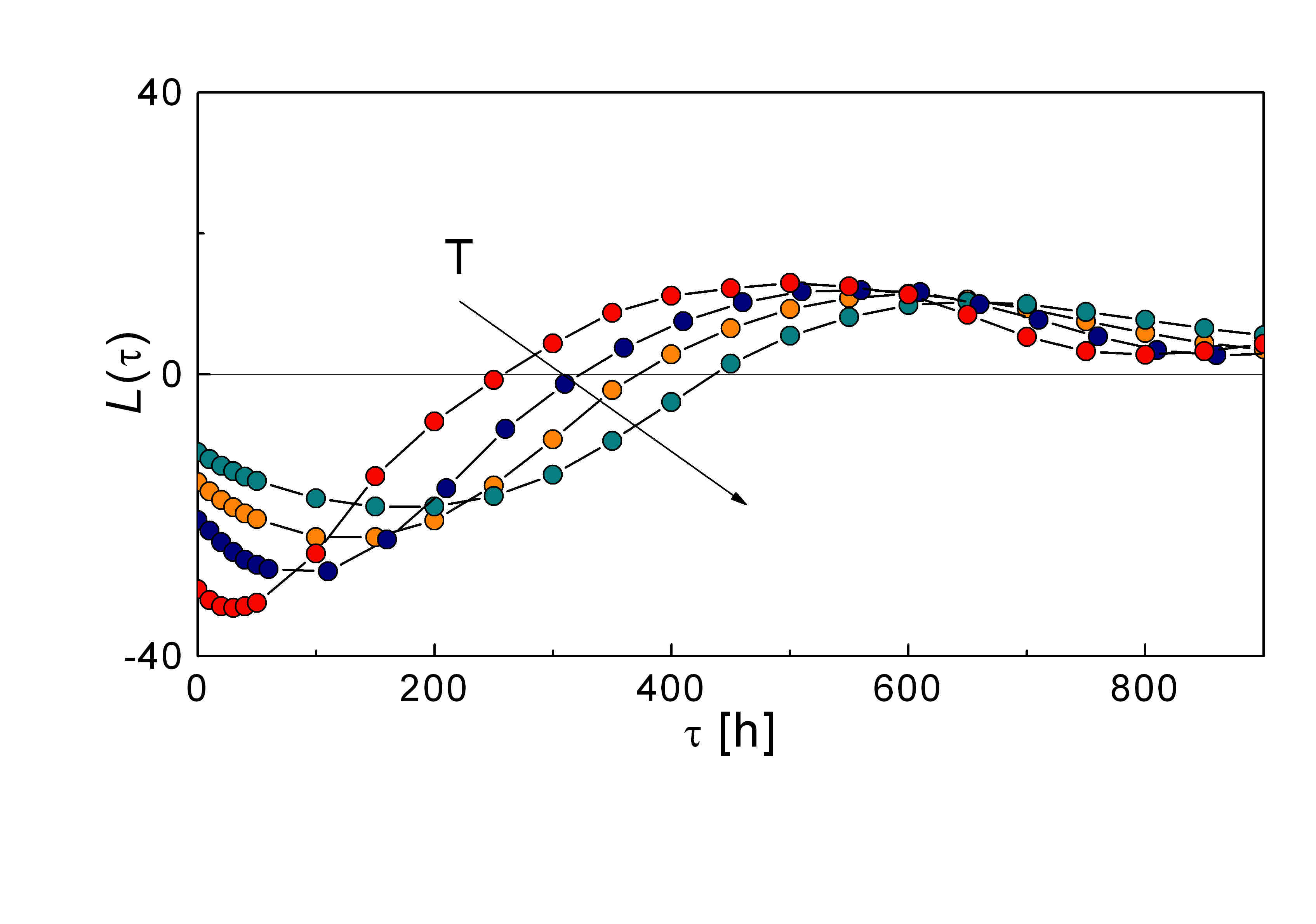}
\caption{\label{fig:figure5}  Leverage function with  volatility windows $T$ = $100h$, $300h$, $660h$, $1000h$. The value of $n$ is $400$ equal for all the curves.}
\end{center}
\end{figure}

\noindent
The function $C_{xy}(\tau)$  for $x=r(t)$ and
$y=\sigma_T(t+\tau)^2$, is shown in Fig.~\ref{fig:figure3}(c).
 The cross-correlation takes negative values at small
$\tau$
  and reaches the minimum   at about
 10-12 days. This indicates that the volatility
increases with negative returns (i.e. with price drops). Then
$C_{xy}(\tau)$
 changes sign relaxing asymptotically to
 zero
 from positive values at large $\tau$. The positive values of $C_{xy}(\tau)$  indicate that the volatility decreases when the returns become positive (i.e. when price rises) and are related to the restored equilibrium within the market (\emph{positive rebound days}).
 It is worthy of remark that the (positive) maximum of the cross-correlation is always smaller than the (negative) minimum. This is the stylized fact known as \emph{down market effect}.
A relevant feature exhibited by the curves in
Figs.~\ref{fig:figure3}~(a-c) is that the zeroes and the extremes of
$C_{xy}(\tau)$ occur at the same values of $\tau$, which is consistent with wide-sense-stationarity for all the  values of $n$.  A further check of wide sense stationarity is provided by the plot of the function $C_{xy}(\tau)n^{-(H_1+H_2)}$. In Fig.~\ref{fig:figure4}, $C_{xy}(\tau)n^{-(H_1+H_2)}$ is plotted with  $x=r(t)$ and $y=\sigma_T(t)$   with
$T=300h$,  $H_1=0.5$ and $H_2=0.77$, $n$ ranges from 100 to 500 with step 100. One can note that the five curves collapse in accord with the invariance of the  product $C_{xy}(\tau) n^{-(H_1+H_2)}$ with $n$.
\par In Fig.~\ref{fig:figure5},
the leverage correlation function
$\mathcal{L}(\tau)=\langle \sigma_T(t+\tau)^2 r(t) \rangle/
\langle r(t)^2\rangle^2$ according to the definition put forward in \cite{Bouchaud}, is plotted for different volatility windows $T$. The function $\langle \sigma_T(t+\tau)^2 r(t) \rangle$ has been calculated by means of Eq.(\ref{crosscovariance}). The negative values of  cross-correlation (at smaller $\tau$) and the following values (\emph{positive rebound days}) at larger $\tau$ can be clearly observed
for several volatility windows $T$. The function $\mathcal{L}(\tau)$
for the DAX stock index, estimated by means of the standard
cross-correlation function, is shown in Figs.~1,2 of
Ref.~\cite{Qiu}. By comparing the curves shown in
Fig.~\ref{fig:figure5} to those of
Ref.~\cite{Qiu}, one can note the higher resolution related to the possibility to detect the correlation at smaller lags (note the  $\tau$ unit is  hours, while in Ref.\cite{Bouchaud,Perello,Qiu,Ahlgren}  is days) and at varying windows $n$, implying the possibility to estimate the degree of cross-correlation at different frequencies. As a final remark, we mention that the cross correlation function between a fractional Brownian motion and its own width can be computed analytically in the large $n$ limit, following the derivation in the Appendix for two general fBm's. The width of a fBm is one possible definition for the volatility, therefore the derivation in the Appendix provides a straightforward estimate of the leverage function.

\subsection{Genomic Sequences}
\begin{figure}
\begin{center}
\includegraphics[width=8cm]{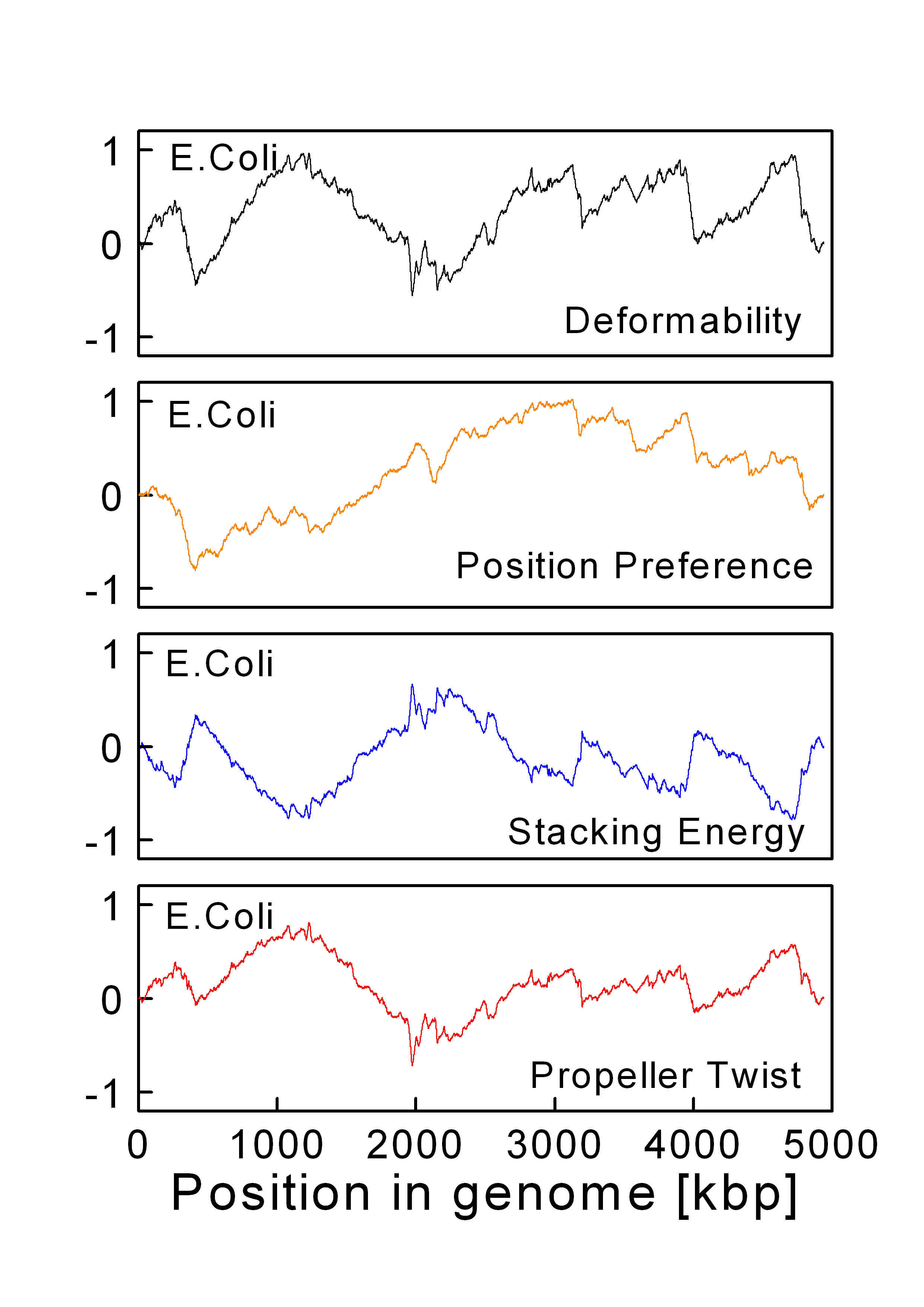}
\caption{\label{fig:figure6}  Structural sequences of
the Escherichia Coli chromosome.}
\end{center}
\end{figure}

 Several  studies are being
addressed to quantify cross-correlations among nucleotide position, intrinsic curvature and flexibility of the
DNA helix, that may ultimately shed light on biological processes, such as protein targeting and transcriptional regulation \cite{Moukhtar,Allen,Pedersen}.
 One problem to
overcome is the comparison of DNA fragments with di- and
trinucleotide scales, hence the need of using high-precision numerical techniques. We consider deformability, stacking energy,
 propeller twist and position preference sequences of the Escherichia
Coli chromosome. The sequences, with details about the
methods used to synthetize/measure the structural properties, are
available at the CBS database - Center for Biological Sequence
Analysis of the Technical University of Denmark
(\textcolor[rgb]{0.00,0.00,1.00}{http://www.cbs.dtu.dk/services/genomeAtlas/}). In order to apply
the proposed method, the average value is subtracted from the
data, that are subsequently integrated to obtain the paths shown
in Fig.~\ref{fig:figure6}. The series are $4938919 bp$ long and have
Hurst exponents: $H=0.70$ (deformability), $H=0.65$ (position preference), $H=0.73$
 (stacking energy), $H=0.70$
 (propeller twist).

\begin{figure}
\begin{center}
\includegraphics[width=8cm]{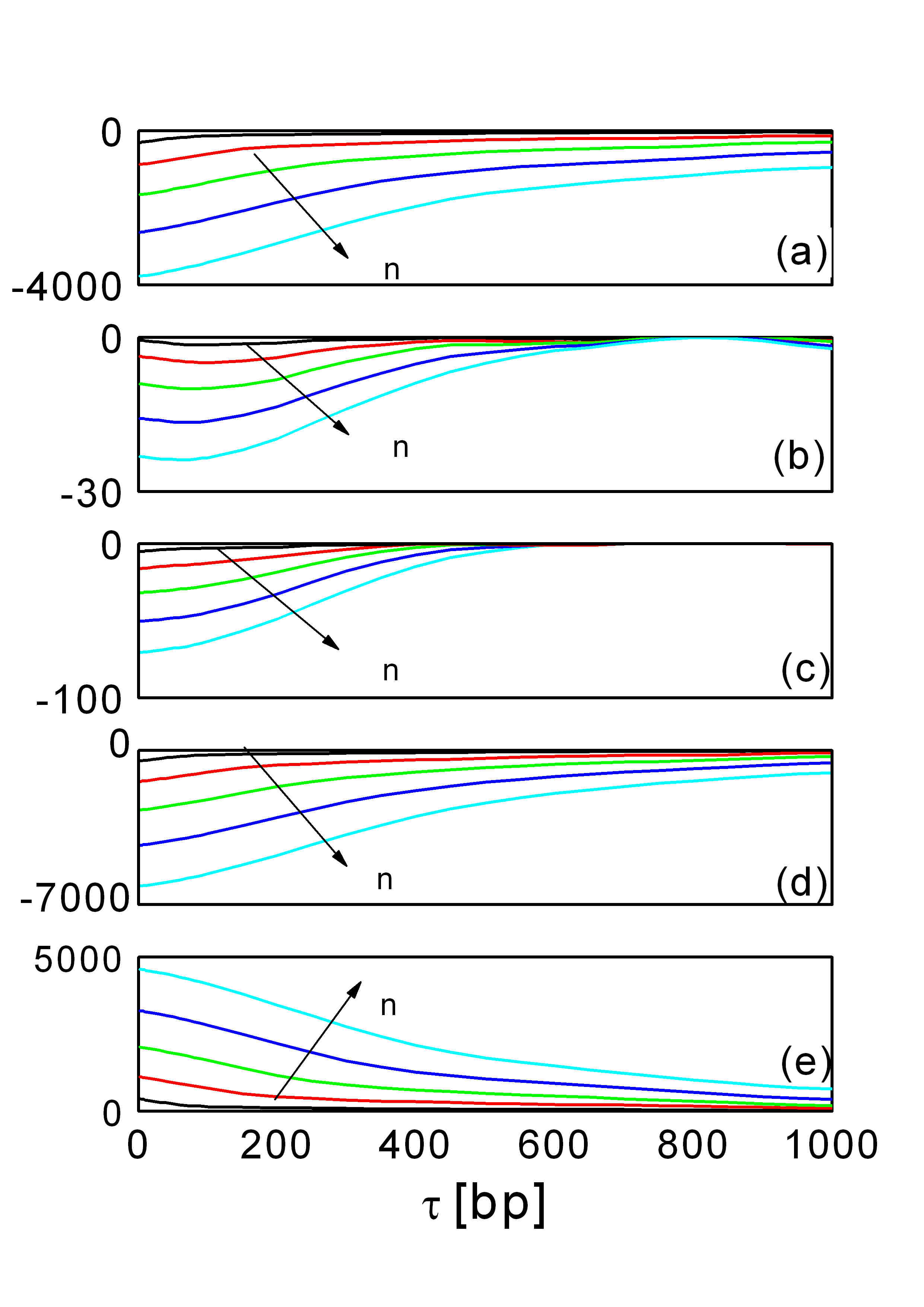}
\caption{\label{fig:figure7} Cross-correlation $C_{xy}(\tau)$ between (a) deformability and
stacking energy; (b) position preference and deformability (c)
propeller twist and position preference; (d) propeller twist and
stacking energy; (e) propeller twist and deformability.  $n$ ranges from 100 to 500 with step
100.}
\end{center}
\end{figure}

\begin{figure}
\begin{center}
\includegraphics[width=8cm]{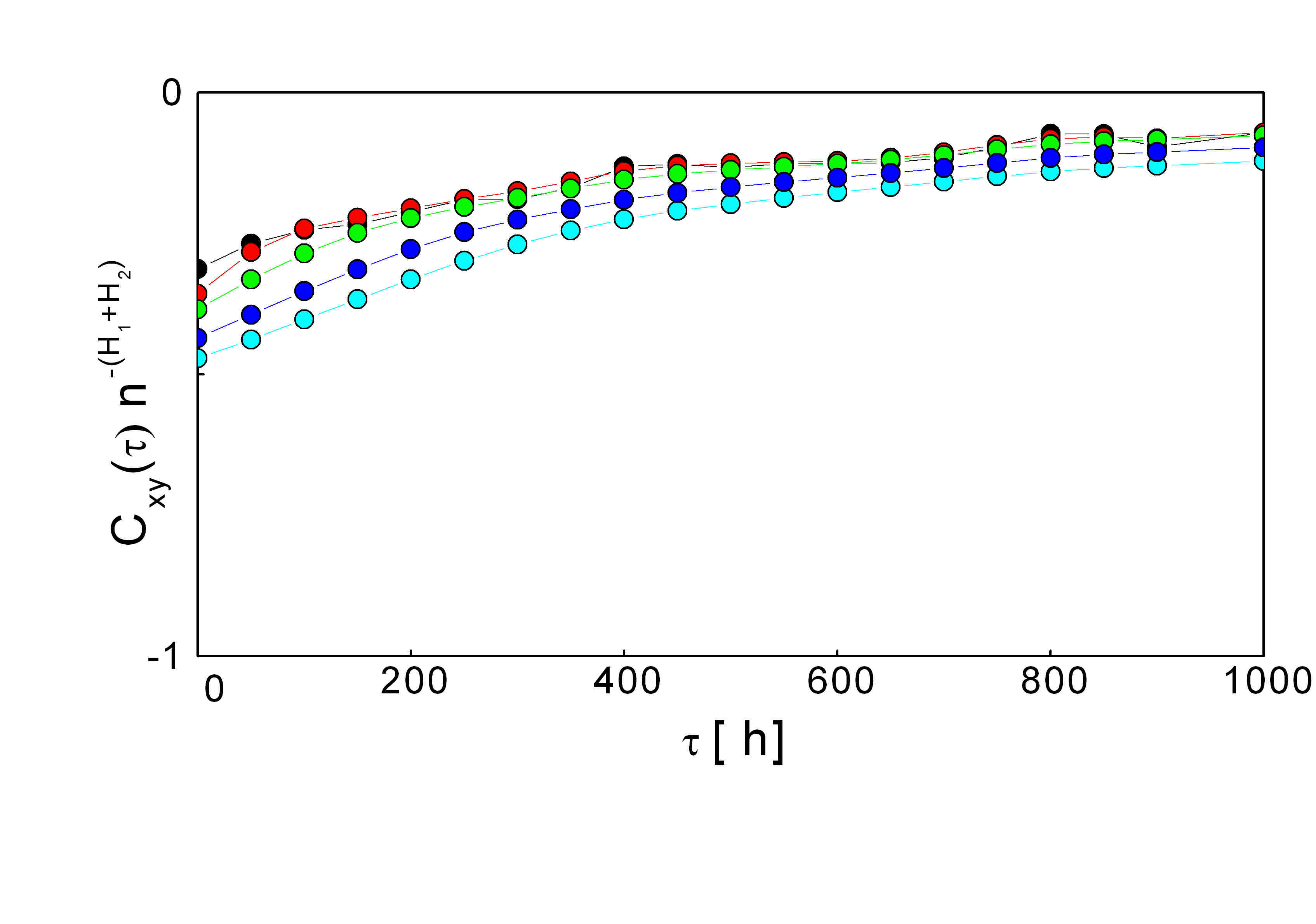}
\caption{\label{fig:figure8} Plot of the function
$C_{xy}(\tau) n^{-(H_1+H_2)}$ with $x(t)$ the deformability, $y(t)$ the
stacking energy, $H_1=0.7$ and $H_2=0.73$.  $n$ ranges from 100 to 500 with step
100. One can note that the five curves collapse,  within the numerical errors of the parameters entering the auto- and cross-corerlation functions. This is in in accord with the invariance of the  product $C_{xy}(\tau) n^{-(H_1+H_2)}$ with the window $n$.}
\end{center}
\end{figure}

The cross-correlation functions $C_{xy}(\tau)$ between
deformability, stacking energy,
 propeller twist and position preference are shown in Fig.~\ref{fig:figure7}~(a-e).
 There is in general a remarkable cross-correlation along the DNA chain indicating the
 existence of interrelated patches of the structural and compositional
 parameters. The  high correlation level between DNA
 flexibility measures and protein complexes indicates that the
 conformation adopted by the DNA bound to a protein depends on the inherent structural features of the DNA.
  It is worthy to remark that the present method provides the dependence of the coupling along the DNA chain rather than simply
  the values of the linear correlation coefficient  $r$. In Table 4 of Ref.~\cite{Pedersen}  one can find the following values of the correlation obtained by either numerical analysis or experimental measurements (in parentheses) over DNA fragments : (a) $r=-0.80~(-0.86)$; (b) $r=0.06~(0.00)$; (c) $r=-0.15~(-0.22)$; (d) $r=-0.74~(-0.82)$;
     (e) $r=-0.80~(-0.87)$.
     Moreover, also for the genomic sequences the function
$C_{xy}(\tau) n^{-(H_1+H_2)}$  is independent of $n$  within the numerical errors of the parameters entering the auto- and cross-correlation functions.  In Fig.~\ref{fig:figure8}, $C_{xy}(\tau) n^{-(H_1+H_2)}$ is shown for  $x(t)$ the deformability, $y(t)$ the
stacking energy, $H_1=0.7$ and $H_2=0.73$.  $n$ ranges from 100 to 500 with step
100.

\section{Conclusions}
 A high-resolution, lag-dependent non-parametric technique based on  Eqs.~(\ref{crosscovariance}-\ref{ytil})
 to measure cross-correlation in long range-correlated series has been developed.
 The  technique has been implemented on (\emph{i}) financial returns and volatilities and
 (\emph{ii})  structural properties of genomic sequences \cite{note}. The results clearly show the existence of coupling regimes characterized
 by positive-negative feedback between the systems at different lags $\tau$ and windows $n$. We point out that - in principle -  other
 methods might be generalized in order to yield estimates of
 the
 cross-correlation between long-range correlated series at varying $\tau$ and $n$.
 However, techniques operating over the series by means of a box division, such as DFA and R/S method,
 are \emph{a-priori} excluded. The box division  causes
 discontinuities in the sliding product of the two series at the extremes of each box,
 and ultimately incorrect estimates of the cross-correlation.
 The present method is not affected by this drawback, since  Eqs.~(\ref{crosscovariance}-\ref{ytil})
 do not require a box division.

\appendix
\section{Details of the calculation:}
Let us start  from Eq.~(\ref{dcaB0}):

\begin{eqnarray}
\label{dcaB0_A} C_{xy}(t,\,\tau)
=\Big\langle\big[B_{H_1}(t)-\widetilde{B}_{H_1}(t)\big]\big[B_{H_2}^*(t+\tau)-\widetilde{B}_{H_2}^*(t+\tau)\big]\Big\rangle
\;\;\;,
\end{eqnarray}
that, after multiplying the terms in parentheses,  becomes:

\begin{eqnarray} \label{dca}
\nonumber
C_{xy}(t,\,t+\tau) &=\Big\langle[B_{H_1}(t)B_{H_2}^*(t+\tau)-B_{H_1}(t)\widetilde{B}_{H_2}^*(t+\tau) \\
& -\widetilde{B}_{H_1}(t)B_{H_2}^*(t+\tau)+\widetilde{B}_{H_1}(t)\widetilde{B}_{H_2}^*(t+\tau)]\Big\rangle \;\;\;.
\end{eqnarray}

In general, the moving average  may be referred to
any point of the moving window, a feature expressed by replacing
Eqs.~(\ref{xtil},\,\ref{ytil})  with
\begin{equation}
\tilde{x}_n(t)= \frac{1}{n}\sum_{k=-\theta n}^{n-\theta n}x(t-k) \label{tx} \qquad \quad
\tilde{y}_n(t+\tau)=\frac{1}{n}\sum_{k=-\theta n}^{n-\theta n}y(t+\tau-k) \label{ty}
\end{equation}
with $0\le\theta\le 1$. In the limit of $n\to\infty$, the sums can be replaced by integrals, so that:
\begin{equation} \label{txy}
\tilde{x}(t)=\int_{-\theta}^{1-\theta}x(\hat{t}-\hat{k}) \qquad \quad \tilde{y}(t+\tau)=\int_{-\theta}^{1-\theta}y(\hat{t}+\hat{\tau}-\hat{k})
\end{equation}
where $t=n\hat{t}$, $ \tau=n\hat{\tau}$, $k=n\hat{k}$.
For the sake of simplicity, the analytical
derivation will be done by using the harmonizable representation
of the fractional Brownian motion~\cite{Benassi,Cohen,Dobric}:
\begin{equation}
\label{harmo}
B_H(t)\equiv\int_{-\infty}^{+\infty}\frac{e^{it\xi}-1}{|\xi|^{H+\frac{1}{2}}}d\bar{B}(\xi)\;,
\end{equation}
where $d\bar{B}(\xi)$ is a representation of $dB(t)$ in the $\xi$
domain. In the following we will consider the case of $t>0$ and $t+\tau>0$.
By using  Eq.~(\ref{harmo}), the cross-correlation
 of two fbms $B_{H_1}(t)$ and $B_{H_2}(t+\tau)$ can be
written as:
\begin{equation} \label{xy}
\hspace{-15mm}\langle B_{H_1}(t)B_{H_2}^*(t+\tau)\rangle=\Big\langle\int_{-\infty}^{+\infty}\frac{e^{it\xi}-1}{|\xi|^{H_1+\frac{1}{2}}}\,d\bar{B}(\xi)
\,\int_{-\infty}^{+\infty}\frac{e^{-i(t+\tau)\eta}-1}{|\eta|^{H_2+\frac{1}{2}}}\,d\bar{B}(\eta)\Big\rangle\;.
\end{equation} Since $d\bar{B}$ is Gaussian, the
following property holds for any $f,\,g\,\in\,L^2(\mathbb{R})$~:
\begin{equation} \label{gaussian}
\Big\langle\int_{-\infty}^{+\infty}f(\xi)d\bar{B}(\xi)\,\left(\int_{-\infty}^{+\infty}g(\eta)d\bar{B}(\eta)\right)^*\Big\rangle=\int_{-\infty}^{+\infty}f(\xi)g^*(\xi)\,d\xi
\end{equation}
By using  Eq.~(\ref{gaussian}), after some
algebra Eq.~(\ref{xy}) writes:
\begin{equation} \label{teo2}
\hspace{-15mm} \langle B_{H_1}(t)B_{H_2}^*(t+\tau)\rangle =D_{H_1,\,H_2}\Big(t^{H_1+H_2}+(t+\tau)^{H_1+H_2}-|\tau|^{H_1+H_2}\Big)\;,\\
\end{equation}
where $D_{H_1,\,H_2}$ is a normalization factor which depends on $H_1$ and $H_2$. In the harmonizable representation of fBm,  $D_{H_1,\,H_2}$ takes the following form \cite{Ayache}:
\begin{equation}
D_{H_1,\,H_2}=D_{H_1+H_2}=-\frac{2}{\pi}\cos\left[\frac{(H_1+H_2)\pi}{2}\right]\Gamma[-(H_1+H_2)]
\end{equation}
normalized such that $D_{H_1,\,H_2}=1$ when $H_1=H_2=\frac{1}{2}$. Different representations of the fBm lead to different values of the coefficient $D_{H_1,\,H_2}$ \cite{Dobric,Stoev}.

\par Eq.~(\ref{teo2}) can be used to calculate each of
the four terms in the right hand side of  Eq.~(\ref{dca}).
 The mean value
of each term in Eq.~(\ref{dca}) is obtained from the general
formula in Eq.~(\ref{teo2}); thus, substituting the right hand
side of Eq.~(\ref{teo2}) and Eq.~(\ref{txy}) into each term in
Eq.~(\ref{dca}) we obtain:
\begin{eqnarray}
\label{dca1}
\hspace{-25mm}&C_{xy} (\hat{t},\,\hat{\tau},\, \theta)=D_{H_1,\,H_2}n^{H_1+H_2}\Big[\Big(\hat{t}^{H_1+H_2}+(\hat{t}+\hat{\tau})^{H_1+H_2}-|\hat{\tau}|^{H_1+H_2}\Big)\nonumber \\
\hspace{-25mm}& -\Big(\hat{t}^{H_1+H_2}+\int_{\hat{h}=-\theta}^{1-\theta}|\hat{t}-\hat{h}+\hat{\tau}|^{H_1+H_2}d\hat{h}-\int_{\hat{h}=-\theta}^{1-\theta}|\hat{t}-\hat{h}|^{H_1+H_2}d\hat{h}\Big)\nonumber \\
\hspace{-25mm}& -\Big(\int_{\hat{k}=-\theta}^{1-\theta}|\hat{t}-\hat{k}|^{H_1+H_2}d\hat{k}+(\hat{t}+\hat{\tau})^{H_1+H_2}-\int_{\hat{k}=-\theta}^{1-\theta}|\hat{t}+\hat{k}|^{H_1+H_2}d\hat{k}\Big) \nonumber \\
\hspace{-25mm}& +\Big(\int_{\hat{k}=-\theta}^{1-\theta}|\hat{t}-\hat{k}|^{H_1+H_2}d\hat{k}+\int_{\hat{h}=-\theta}^{1-\theta}|\hat{t}-\hat{h}+\hat{\tau}|^{H_1+H_2}d\hat{h}\nonumber \\
\hspace{-25mm}& - \int_{\hat{h}=-\theta}^{1-\theta}\int_{\hat{k}=-\theta}^{1-\theta }|\hat{\tau}-\hat{h}-\hat{k}|^{H_1+H_2}d\hat{h}\,d\hat{k}\Big)\Big]
\end{eqnarray}
where each term in round parentheses corresponds to each of
the four terms in Eq.~(\ref{dca}). Summing the terms in
Eq.~(\ref{dca1}), one can notice that  time
$t$ cancels out, thus one finally obtains:
\begin{eqnarray}
\label{integral}
 \hspace{-25mm}C_{xy}(\hat{\tau},\,\theta) &= n^{H_1+H_2}D_{H_1,\,H_2}\Big[-\hat{\tau}^{H_1+H_2}+\int_{-\theta}^{1-\theta}|\hat{\tau}-\hat{h}|^{H_1+H_2}\,d\hat{h}\nonumber \\
 \hspace{-25mm} &+\int_{-\theta}^{1-\theta}|\hat{\tau}+\hat{k}|^{H_1+H_2}\,d\hat{k}
 -\int_{-\theta}^{1-\theta}|\hat{\tau}-\hat{h}+\hat{k}|^{H_1+H_2}\,d\hat{h}\,d\hat{k}\;\Big]\;, \nonumber \\\hspace{-25mm}&
 \end{eqnarray}

Consistently with the large $n$ limit, we take $\tau<n$, namely $\hat{\tau}<1$. The integral (\ref{integral}) admits four different solutions, depending on the values taken by the parameters $\hat{\tau}$ and $\theta$. Let us consider each case separately.

\vskip 1cm

\paragraph*{Case 1: $\hat{\tau}<\theta$ and $\hat{\tau}+\theta<1$}

\begin{eqnarray}
\label{case1}
\nonumber
\hspace{-25mm}C_{xy}(\hat{\tau},\,\theta)&= n^{H_1+H_2}D_{H_1,\,H_2}\Big[-\hat{\tau}^{H_1+H_2}-\frac{(1-\hat{\tau})^{2+H_1+H_2}-2\hat{\tau}^{2+H_1+H_2}+(1+\hat{\tau})^{2+H_1+H_2}}{(1+H_1+H_2)(2+H_1+H_2)} \nonumber\\
\hspace{-25mm}& \nonumber \\ \nonumber
\hspace{-25mm}&+\frac{(1+\hat{\tau}-\theta)^{1+H_1+H_2}+(\theta-\hat{\tau})^{1+H_1+H_2}+(1-\hat{\tau}-\theta)^{1+H_1+H_2}+(\hat{\tau}+\theta)^{1+H_1+H_2}}{1+H_1+H_2}\Big]
\nonumber \\\hspace{-25mm}&
\end{eqnarray}

\paragraph*{Case 2: $\hat{\tau}<\theta$ and $\hat{\tau}+\theta>1$}

\begin{eqnarray}
\label{case2}
\nonumber
\hspace{-25mm}C_{xy}(\hat{\tau},\,\theta)&= n^{H_1+H_2}D_{H_1,\,H_2}\Big[-\hat{\tau}^{H_1+H_2}-\frac{(1-\hat{\tau})^{2+H_1+H_2}-2\hat{\tau}^{2+H_1+H_2}+(1+\hat{\tau})^{2+H_1+H_2}}{(1+H_1+H_2)(2+H_1+H_2)} \nonumber\\
\hspace{-25mm}& \nonumber \\ \nonumber
\hspace{-25mm}&+\frac{(1+\hat{\tau}-\theta)^{1+H_1+H_2}+(\theta-\hat{\tau})^{1+H_1+H_2}-(\hat{\tau}+\theta-1)^{1+H_1+H_2}+(\hat{\tau}+\theta)^{1+H_1+H_2}}{1+H_1+H_2}\Big]
\nonumber \\\hspace{-25mm}&
\end{eqnarray}

\paragraph*{Case 3: $\hat{\tau}>\theta$ and $\hat{\tau}+\theta<1$}
\begin{eqnarray}
\label{case3}
\nonumber
\hspace{-25mm}C_{xy}(\hat{\tau},\,\theta)&= n^{H_1+H_2}D_{H_1,\,H_2}\Big[-\hat{\tau}^{H_1+H_2}-\frac{(1-\hat{\tau})^{2+H_1+H_2}-2\hat{\tau}^{2+H_1+H_2}+(1+\hat{\tau})^{2+H_1+H_2}}{(1+H_1+H_2)(2+H_1+H_2)} \nonumber\\
\hspace{-25mm}& \nonumber \\
\nonumber
\hspace{-25mm}&+\frac{(1+\hat{\tau}-\theta)^{1+H_1+H_2}-(\hat{\tau}-\theta)^{1+H_1+H_2}+(1-\hat{\tau}-\theta)^{1+H_1+H_2}+(\hat{\tau}+\theta)^{1+H_1+H_2}}{1+H_1+H_2}\Big]
\nonumber \\\hspace{-25mm}&
\end{eqnarray}
 It is easy to see that this case includes the Eq.~(2.5) treated
in the paper.

\paragraph*{Case 4: $\hat{\tau}>\theta$ and $\hat{\tau}+\theta>1$}
\begin{eqnarray}
\label{case4}
\nonumber
\hspace{-25mm}C_{xy}(\hat{\tau},\,\theta)&= n^{H_1+H_2}D_{H_1,\,H_2}\Big[-\hat{\tau}^{H_1+H_2}-\frac{(1-\hat{\tau})^{2+H_1+H_2}-2\hat{\tau}^{2+H_1+H_2}+(1+\hat{\tau})^{2+H_1+H_2}}{(1+H_1+H_2)(2+H_1+H_2)} \nonumber\\
\hspace{-25mm}& \nonumber \\ \nonumber
\hspace{-25mm}&+\frac{(1+\hat{\tau}-\theta)^{1+H_1+H_2}-(\hat{\tau}-\theta)^{1+H_1+H_2}-(\hat{\tau}+\theta-1)^{1+H_1+H_2}+(\hat{\tau}+\theta)^{1+H_1+H_2}}{1+H_1+H_2}\Big]
\nonumber \\\hspace{-25mm}&
\end{eqnarray}


\section*{References}
\begin{thebibliography}{10}
\bibitem{Rosenblum} M.~Rosenblum and A.~Pikovsky,  (2007) Phys.~Rev.~Lett.~{\bf 98}, 064101.
\bibitem{Zhou} T.~Zhou, L.~Chen and K.~Aihara, (2005) Phys.~Rev.~Lett.~{\bf 95}, 178103.
\bibitem{Oberholzer} S.~Oberholzer et al.  (2006) Phys.~Rev.~Lett.~ {\bf 96},  046804.
\bibitem{Dhamala} M.~Dhamala, G.~Rangarajan, (2008) M.~Ding, Phys.~Rev.~Lett.~ {\bf 100}, 018701.
\bibitem{Verdes} P.~F.~Verdes, (2005) Phys.~Rev.~E {\bf 72}, 026222.
\bibitem{Palus} M.~Palus and M.~Vejmelka,  (2007) Phys.~Rev.~E {\bf 75}, 056211.
\bibitem{Kreuz} T.~Kreuz et al. (2007) Physica D  {\bf 225}, 29 .
\bibitem{Du} Lu-Chun Du and Dong-Cheng Mei, (2008) J. Stat. Mech.  P11020.
\bibitem{Tass} P.~Tass et al. (1998) Phys.~Rev.~Lett.~{\bf 81}, 3291.
\bibitem{Huybers} P.~Huybers,  (2006) W.~Curry, Nature {\bf 441}, 7091.
\bibitem{Ashkenazy} Y.~Ashkenazy, (2006)  Climate Dynamics {\bf 27}, 421.
\bibitem{Black} F.~Black, (1976) J.~of~Fin.~Econ. {\bf 3}, 167.
\bibitem{Schwert}W. Schwert, J.~of~Finance   (1989) {\bf 44}, 1115.
\bibitem{Haugen}R.~Haugen, E.~Talmor, W.~Torous, (1991)  J.~of~Finance  {\bf 44}, 1115.
\bibitem{Glosten} L. Glosten, J. Ravi  and D. Runkle,  (1992)  J.~of~Finance {\bf 48}, 1779.
\bibitem{Wu1} G.~Bekaert, G.~Wu, (2000) The Review of Financial Studies {\bf 13}, 1.
\bibitem{Figlewski} S. Figlewski,   X. Wang  (2000) \emph{Is the 'Leverage Effect' a Leverage Effect? }, Working Paper, Stern School of Business, New York.
\bibitem{Bouchaud} J.~P.~Bouchaud, A.~Matacz and M.~Potters,  (2001) Phys.~Rev.~Lett. {\bf 87}, 228701.
\bibitem{Perello} J.~Perello and J.~Masoliver, (2003) Phys.~Rev.~E {\bf 67}, 037102.
\bibitem{Qiu} T.~Qiu, B.~Zheng, F.~Ren and S.~Trimper,  (2006) Phys.~Rev.~E {\bf 73}, 065103(R).
\bibitem{Ahlgren} R.~Donangelo, M.~H.~Jensen, I.~Simonsen,  K.~Sneppen, (2006) J.~Stat.~Mech.~ L11001.
\bibitem{Varga} I Varga-Haszonits and I Kondor, (2008) J. Stat. Mech.  P12007.
\bibitem{Montero} M. Montero,  (2007) J.~Stat.~Mech. P04002.
\bibitem{Moukhtar} J.~Moukhtar, E.~Fontaine, C.~Faivre-Moskalenko and A.~Arneodo, (2007)
 Phys.~Rev.~Lett. {\bf 98}, 178101.
\bibitem{Allen} T.E.~Allen, N.D.~Price, A.~Joyce and B.O.~Palsson, (2006) PLoS Computational Biology, {\bf 2}, e2.
\bibitem{Pedersen} A.G.~Pedersen, L.J.~Jensen, S.~Brunk, H.H.~
Staerfeld and D.W.~Ussery, (2000) J.~Mol.~Biol. {\bf 299}, 907.
\bibitem{Jun}  W.C.~Jun,  G.~Oh and S.~Kim, (2006)  Phys.~Rev.~E {\bf 73}, 066128.
 \bibitem{Podobnik} B.~Podobnik, H.E.~Stanley, (2008) Phys.~Rev.~Lett. {\bf 100}, 084102.
\bibitem{Mandelbrot}
B.~B.~Mandelbrot, J.~W.~Van Ness,  (1968) SIAM Rev. {\bf 4}, 422.
\bibitem{Carbone1} A.~Carbone, G.~Castelli,  H.~E.~Stanley,
 Phys.~Rev.~E {\bf 69}, 026105 (2004).
 \bibitem{Carbone2}  A.~Carbone, Phys.~Rev.~E {\bf 76}, 056703
  (2007).
    \bibitem{Carbone3} S.~Arianos and A.~Carbone,
  Physica A {\bf 382}, 9 (2007).
  \bibitem{Carbone4}  A.~Carbone and H.~E.~Stanley,
  Physica A {\bf 384}, 21 (2007).
  \bibitem{Carbone5}  A.~Carbone and H.~E.~Stanley,
  Physica A {\bf 340}, 544 (2004).
\bibitem{note} The MATLAB and C++ codes implementing the proposed method, the DAX and E-COLI sequences used in this work are downloadable at: \textcolor[rgb]{0.00,0.00,1.00}{www.polito.it/noiselab/utilities}

\bibitem{Benassi}
A.~Benassi, S.~Jaffard, D.~Roux, Rev.~Mat.~Iber. {\bf 13}, 19,
(1997).

\bibitem{Cohen}
S.~Cohen,  Fractals: Theory and Applications in Engineering.
M.~Dekking, J.~L\'evy V\'ehel, E.~Lutton and C.~Tricot (Eds.).
Springer Verlag, 1999.

\bibitem{Ayache}
A.~Ayache, S.~Cohen, J.~Levy Vehel, Proceedings of the conference ICASSP, Istanbul June 2000.

\bibitem{Dobric}
V.~Dobric, F.~M.~Ojeda,  IMS Lecture Notes-Monograph Series,
\emph{High Dimensional Probability}, {\bf 51}, 77, (2006).

\bibitem{Stoev}
S.~A.~Stoev, M.~S.~Taqqu, Stochastic Processes and their Applications {\bf 116}, 200 (2006).

\end{thebibliography}
\end{document}